\documentclass[journal,comsoc,twocolumn]{IEEEtran}
\usepackage[utf8]{inputenc}
\usepackage[T1]{fontenc}
\usepackage{float}
\usepackage{amsmath}
\usepackage{amsfonts}
\usepackage{mathtools}
\usepackage{amsfonts}
\IEEEoverridecommandlockouts
\usepackage{amssymb}
\usepackage{graphicx}
\usepackage{epstopdf}
\usepackage{makeidx}
\usepackage{csquotes}
\usepackage{graphicx}
\usepackage{cite}
\usepackage{xcolor}
\usepackage{subcaption}
\usepackage{graphicx}
\usepackage{balance}
\usepackage{pifont}
\begin{document}
%
\title{Beyond Diagonal RIS: A New Frontier for 6G Internet of Things Networks}

\author{Wali Ullah Khan, Chandan~Kumar~Sheemar, Eva Lagunas, Symeon Chatzinotas,~\IEEEmembership{Fellow,~IEEE,} \thanks{Wali Ullah Khan, Chandan Kumar Sheemar, Eva Lagunas, and Symeon Chatzinotas are with the the Interdisciplinary Centre for Security, Reliability, and Trust (SnT), University of Luxembourg, 1855 Luxembourg City, Luxembourg (e-mails: \{waliullah.khan, chandankumar.sheemar, eva.lagunas, symeon.chatzinotas\}@uni.lu)

}}%

\markboth{IEEE Magazine (For Review)
}
{Shell \MakeLowercase{\textit{et al.}}: Bare Demo of IEEEtran.cls for IEEE Journals} 

\maketitle

\begin{abstract}
Reconfigurable intelligent surface (RIS) technology has emerged as a promising enabler for next-generation wireless networks, offering a paradigm shift from passive environments to programmable radio wave propagation. Despite the potential of diagonal RIS (D-RIS), its limited wave manipulation capability restricts performance gains. In this paper, we investigate the burgeoning concept of beyond-diagonal RIS (BD-RIS), which incorporates non-diagonal elements in its scattering matrix to deliver more fine-grained control of electromagnetic wavefronts. We begin by discussing the limitations of traditional D-RIS and introduce key BD-RIS architectures with different operating modes. We then highlight the features that make BD-RIS particularly advantageous for 6G IoT applications, including advanced beamforming, enhanced interference mitigation, and flexible coverage. A case study on BD-RIS-assisted vehicle-to-vehicle (V2V) communication in an underlay cellular network demonstrates considerable improvements in spectral efficiency when compared to D-RIS and conventional systems. Lastly, we present current challenges such as hardware design complexity, channel estimation, and non-ideal hardware effects, and propose future research directions involving AI-driven optimization, joint communication and sensing, and physical layer security. Our findings illustrate the transformative potential of BD-RIS in shaping high-performance, scalable, and reliable 6G IoT networks.
\end{abstract}


\IEEEpeerreviewmaketitle

\section{Introduction}

The next-generation sixth-generation (6G) wireless networks are expected to revolutionize global connectivity by surpassing the limitations of 5G and introducing unprecedented performance enhancements. Envisioned to operate in the early 2030s, 6G will support ultra-high data rates, sub-millisecond latency, massive device connectivity, and intelligent network automation \cite{9040264}. These capabilities are essential for enabling futuristic applications such as autonomous systems, extended reality (XR), real-time holographic communications, and precision healthcare. A fundamental pillar of 6G will be the Internet of Things (IoT), which will interconnect billions of smart devices, driving advancements in smart agriculture, intelligent transportation, industrial automation, and healthcare monitoring \cite{9319211}. However, ensuring reliable and energy-efficient communication in dense IoT deployments remains a major challenge, necessitating innovative approaches to enhance wireless propagation and system capacity.

To address these challenges, Reconfigurable Intelligent Surfaces (RIS) have emerged as a disruptive technology capable of reconfiguring wireless environments dynamically \cite{10584518}. By leveraging passive elements to manipulate the phase of incident electromagnetic waves, RIS can enhance coverage, mitigate interference, and improve spectral efficiency, making it highly suitable for 6G IoT networks. However, conventional RIS architectures are constrained by only diagonal reflection matrix response, limiting their ability to manage complex, dense and dynamic IoT scenarios \cite{10716670}.

A promising evolution of RIS, known as Beyond Diagonal RIS (BD-RIS), has been introduced to overcome these limitations and unlock greater flexibility in signal manipulation \cite{9514409}. Unlike traditional D-RIS, which controls only the phase of incoming signals, BD-RIS incorporates interconnected impedance elements, allowing control over both phase and amplitude, including also the control over off-diagonal elements \cite{9913356}. This enhanced capability enables more efficient multi-user beamforming, dynamic interference suppression, and improved signal reception, which are crucial for supporting IoT networks in diverse 6G environments. Moreover, BD-RIS can optimize energy consumption, making it an attractive solution for power-constrained IoT devices \cite{10784592}. By expanding the functionality of RIS beyond diagonal phase-shift matrices, BD-RIS represents a transformative step toward intelligent and reconfigurable wireless networks, unlocking new possibilities for next-generation IoT communications.

In this article, we have provided a comprehensive analysis of BD-RIS technology for 6G IoT networks, beginning with an overview of RIS fundamentals and the limitations of conventional diagonal RIS. We then introduced BD-RIS as an evolution of RIS design, detailing its architectural classifications—single-connected, fully-connected, and group-connected—and their respective operating modes (reflective, transmissive, hybrid, and multi-sector). We highlighted the unique advantages of BD-RIS, including enhanced beamforming, dynamic adaptability, and superior interference management, making it ideal for emerging IoT applications in sectors such as smart agriculture, industry 5.0, smart cities, and non-terrestrial networks (NTNs). A practical case study on BD-RIS-assisted vehicle-to-vehicle (V2V) communication was presented to demonstrate the significant performance gains in spectral efficiency when BD-RIS is deployed to optimize both direct and indirect paths. Finally, we discussed ongoing challenges—such as hardware design constraints, channel estimation, and real-world impairments—and outlined key research directions, including AI-driven optimization, joint communication and sensing (JCAS), and physical layer security. These findings affirm BD-RIS as a pivotal enabler for next-generation networks, driving forward the vision of ultra-reliable, energy-efficient, and pervasive connectivity across diverse 6G IoT environments

\emph{Paper Organization:}
This paper begins with an overview of RIS fundamentals and the evolution from diagonal RIS to BD-RIS in Section II. In Section III, we present key 6G IoT use cases, followed by a V2V communication case study in Section IV. Section V identifies implementation challenges and possible research directions, while Section VI concludes the work with insights and future prospects.
\section{Preliminaries of RIS Technology}

\subsection{Overview of RIS}
RIS technology has emerged as a transformative solution in wireless communications, offering the potential to revolutionize how electromagnetic waves interact with their environment \cite{8811733}. By intelligently manipulating the propagation of wireless signals, RIS shifts the paradigm from uncontrollable and passive wireless environments to smart, programmable, and adaptive systems. A typical RIS system consists of an array of nearly passive reconfigurable elements, each capable of adjusting the amplitude and phase of incident electromagnetic waves. By dynamically tuning these properties, RIS can enhance signal strength at the receiver, suppress interference, and improve coverage, especially in scenarios where traditional methods fall short \cite{iacovelli2024holographic}. This functionality makes RIS particularly well-suited for next-generation wireless systems, including 6G and beyond, where energy efficiency, spectral efficiency, and ultra-reliable communication are critical.

RIS can generally be categorized into diagonal RIS (D-RIS) and BD-RIS. The D-RIS design relies on a reconfigurable matrix where only the diagonal elements are non-zero. Each element of the RIS independently modifies the amplitude and phase of the incident signal without interacting with other elements, resulting in a reflective beamforming matrix that is diagonal and offers limited beamforming capabilities \cite{10396846}. In D-RIS, each element is equipped with a tunable meta-atom that adjusts the phase shift applied to the incident wave. By coordinating these phase shifts, the RIS steers the signal toward the intended direction, enhancing received power or mitigating interference.

While D-RIS is inexpensive and simple, it has some drawbacks. Its separate element control limits its capacity to shape the wavefront with high accuracy, making it ineffective in complicated environments like dense multipath scenarios. Furthermore, in extremely dynamic or irregular propagation settings, D-RIS may have performance constraints. Its absence of inter-element connections limits its ability to handle multi-user interference and provide fine-grained beam steering. In such circumstances, more advanced designs like BD-RIS provide greater adaptability and efficiency, addressing the limits of D-RIS and driving the investigation of further RIS concepts \cite{10817282}.

\subsection{BD-RIS: The Next Frontier}
BD-RIS represents a significant evolution of RIS technology, addressing the limitations of traditional D-RIS designs  \cite{9913356}. Unlike D-RIS, BD-RIS incorporates non-diagonal elements in its phase shift matrix, enabling a more versatile and comprehensive manipulation of electromagnetic waves \cite{9514409}. This interconnection of reconfigurable elements enables BD-RIS to perform advanced wave manipulations. For example, BD-RIS tailoring the reflected wavefront to achieve desired beam patterns. It modulates the scattered signals to enhance coverage and reduce interference. Moreover, BD-RIS adjusts to varying environmental conditions and user requirements in real time. Along with the three operating modes similar to D-RIS—reflecting, transmissive, and hybrid—BD-RIS introduces an extended hybrid mode called the multi-sector mode, where the surface is divided into more than two sectors \cite{10158988}. This mode achieves superior performance enhancements over the hybrid mode by employing high-gain reconfigurable elements with narrower beamwidths capable of covering a full space. 

\begin{table*}[!t] 
\renewcommand{\arraystretch}{1.2} 
\caption{Different modes and architectures of BD-RIS along with their characteristic and complexity.}
\centering
\begin{tabular}{|c|c|c|c|}
\hline
Characteristics & Single-connected architecture & Fully-connected architecture & Group-connected architecture  \\
\hline\hline
Number of groups & $N$ & 1 & $G$ \\
\hline
Group dimension & 1 & $N$ & $\hat{N}$ \\
\hline
Elements/group & 1 & $N^2$ & $\hat{N}^2$\\
\hline
Number of Non-zero elements & $N$ & $N^2$ & $G\hat{N}^2$ \\
\hline
Reflective mode & $|\phi_{r,n}|^2=1$ & ${\bf \Phi}^H_{r}{\bf \Phi}_{r}={\bf I}_N$ & ${\bf \Phi}^H_{r,g}{\bf \Phi}_{r,g}={\bf I}_{\hat{N}}$  \\
\hline
Transmissive mode & $|\phi_{t,n}|^2=1$  & ${\bf \Phi}^H_{t}{\bf \Phi}_{t}={\bf I}_N$ & ${\bf \Phi}^H_{t,g}{\bf \Phi}_{t,g}={\bf I}_{\hat{N}}$  \\
\hline
Hybrid mode &$|\phi_{r,n}|^2+|\phi_{t,n}|^2=1$ & ${\bf \Phi}^H_{r}{\bf \Phi}_{r}+{\bf \Phi}^H_{t}{\bf \Phi}_{t}={\bf I}_N$  & ${\bf \Phi}^H_{r,g}{\bf \Phi}_{r,g}+{\bf \Phi}^H_{t,g}{\bf \Phi}_{t,g}={\bf I}_{\hat{N}}$ \\
\hline
Hardware complexity  &R$^\star$,T$^\star$: $N$, H$^\star$: $(3/2)N$, M$^\star$: $(S+1)K/2$ & R$^\star$,H$^\star$,M$^\star$: $(N+1)\frac{N}{2}$  & R$^\star$,H$^\star$,M$^\star$: $(\frac{N}{G}+1)\frac{N}{2}$ \\
\hline
\multicolumn{4}{l}{$^{\star}$R-Reflective, T-Transmissive, H-Hybrid, M-Multi sector.}
\end{tabular}   
\label{Table2} 
\end{table*} 

\subsection{Classification of BD-RIS}
This section categorizes BD-RIS architectures and their operational modes.

\subsubsection{BD-RIS Architecture}
\begin{figure}[!t]
\centering
\includegraphics[width=0.48\textwidth]{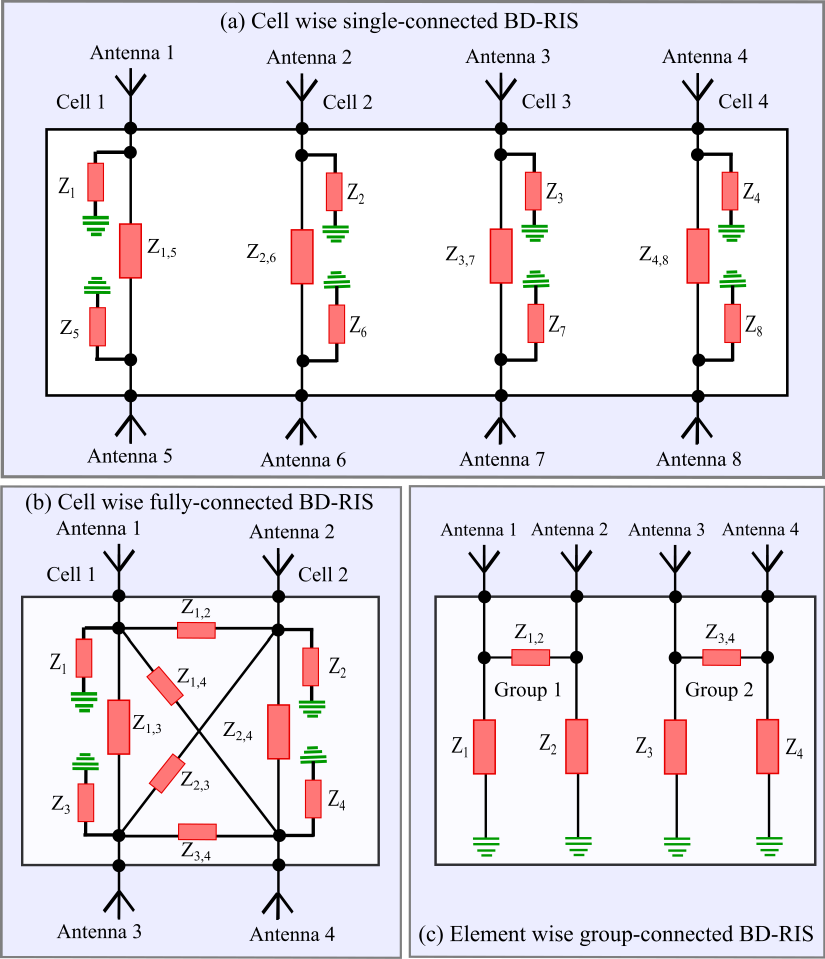}
\caption{Architectures of BD-RIS: (a) cell-wise single-connected BD-RIS with four cells, (b) cell-wise fully-connected BD-RIS with two cells, and (c) element-wise group-connected BD-RIS with two groups.}
\label{ArchitectureBDRIS}
\end{figure}

As illustrated in Fig. \ref{ArchitectureBDRIS}, BD-RIS architectures are classified into three types:  
(a) \textit{Single-connected BD-RIS}, where each element independently controls phase and amplitude without interconnections \cite{10316535}. This structure simplifies implementation but restricts spatial degrees of freedom due to diagonal phase shift matrices.  
(b) \textit{Fully-connected BD-RIS}, which interconnects all elements via an impedance network, forming a non-diagonal phase shift matrix that enables superior beamforming and scattering capabilities \cite{10716670}. Despite its performance benefits, complexity and power consumption increase with the number of elements.  
(c) \textit{Group-connected BD-RIS}, which partitions elements into subarrays, each forming a fully-connected structure \cite{9913356}. This block-diagonal phase shift matrix balances complexity and performance, making it scalable for large deployments.

Fig. \ref{ArchitectureBDRIS} illustrates these architectures:  
(a) Single-connected BD-RIS, where multiple reconfigurable elements in a cell dynamically balance reflection and transmission.  
(b) Fully-connected BD-RIS, where each element influences all others within a cell for precise control.  
(c) Group-connected BD-RIS, where elements are grouped to optimize efficiency while reducing implementation costs.

These architectures represent a trade-off between complexity, scalability, and performance. Single-connected BD-RIS offers simplicity, fully-connected BD-RIS maximizes control at the cost of complexity, and group-connected BD-RIS provides a scalable middle ground.

\begin{figure*}[!t]
\centering
\includegraphics [width=0.9\textwidth]{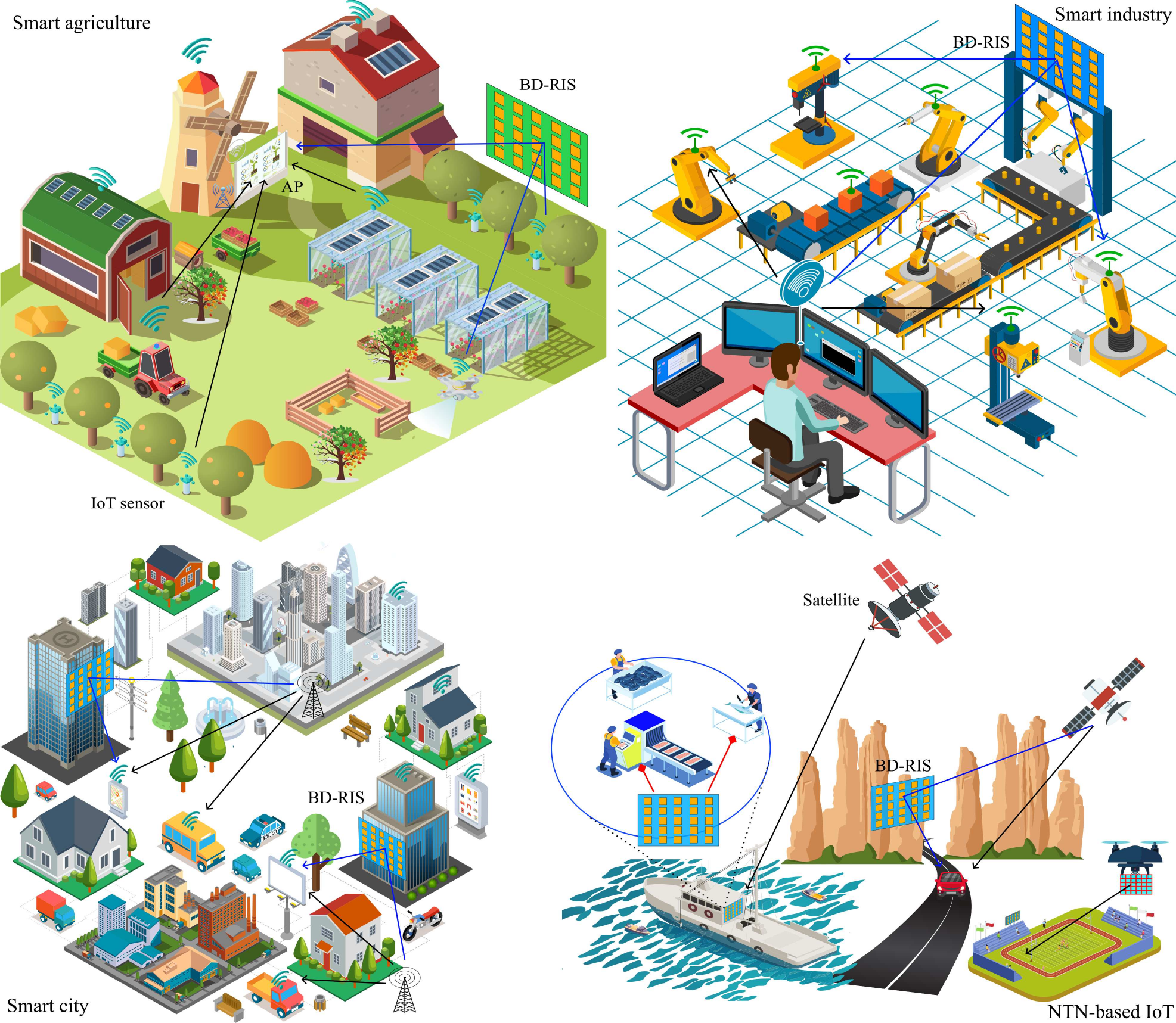}
\caption{Use cases of BD-RIS in 6G IoT networks.}
\label{UseCases}
\end{figure*}
\subsubsection{BD-RIS Operating Modes}
BD-RIS can operate in four modes, i.e., reflective, transmissive, hybrid, and multi-sector modes.
(a) \textit{Reflective mode:} In reflective mode, the transmitter and receiver are on the same side of the BD-RIS, providing half-space coverage. The incident signal is reflected back on the same side, potentially causing self-interference that degrades performance \cite{10158988}. Reflective BD-RIS is suitable for installation on walls, facades, or ceilings in terrestrial setups. (b) \textit{Transmissive mode:}
In transmissive mode, the transmitter and receiver are on opposite sides of the BD-RIS, allowing the incident signal to pass through \cite{10319662}. This mode improves signal penetration through barriers such as walls and windows. While similar to reflective BD-RIS, its ability to refract signals enhances its deployment flexibility in complex environments. (c) \textit{Hybrid mode:} Hybrid BD-RIS combines reflection and transmission for full-space coverage. Known as STAR-RIS in a single-connected architecture, it uses two back-to-back reconfigurable elements with unidirectional radiation patterns for each cell \cite{10133841}. These elements provide reflection and transmission coverage, and are connected through a two-port impedance network. (d) \textit{Multi-sector mode:} Multi-sector BD-RIS extends the hybrid mode by dividing the area into multiple sectors, each served by a dedicated set of reconfigurable elements arranged in an $S$-sided polygon to cover $1/S$ of the area \cite{10316535}. Connected through an $S$-port impedance network, this mode enhances coverage with narrower beamwidth and higher gain compared to the hybrid mode.

These modes offer flexibility and efficiency for 6G IoT networks, ensuring optimized performance and connectivity. A summary of the characteristics and complexity of various BD-RIS architectures and modes is provided in Table \ref{Table2}.

\subsection{Features of BD-RIS in 6G IoT Networks}

BD-RIS presents significant advantages over conventional D-RIS by leveraging inter-element connectivity and a full scattering matrix, making it particularly beneficial for IoT applications in 6G networks.  
\subsubsection{Enhanced Beamforming and Flexibility} Unlike D-RIS, which only adjusts phase shifts, BD-RIS controls both amplitude and phase, enabling precise and adaptive beamforming. This improves signal strength and quality, crucial for IoT networks with diverse connectivity demands.  

\subsubsection{Adaptability to Dynamic IoT Environments} BD-RIS efficiently reconfigures itself to accommodate fluctuating IoT traffic and mobility patterns, ensuring seamless connectivity even in highly dynamic or obstructed environments.  

\subsubsection{Superior Interference Management} By jointly controlling phase and amplitude, BD-RIS effectively mitigates interference in complex IoT deployments, such as smart cities, industrial automation, and NTNs, where multiple wireless technologies coexist.  

\subsubsection{Extended Coverage and Connectivity} BD-RIS supports full-space (360-degree) coverage, making it highly suitable for large-scale IoT applications, including agricultural monitoring, remote sensing, and smart grid networks.  

 \subsubsection{Versatile Applications} The ability to adapt to non-line-of-sight (NLoS) conditions, mobility constraints, and dynamic IoT scenarios makes BD-RIS ideal for use in UAV-assisted IoT, vehicular communications, and energy-efficient smart infrastructure.  

By addressing key limitations of D-RIS, BD-RIS enhances the efficiency, reliability, and scalability of 6G-enabled IoT ecosystems, ensuring robust connectivity for future smart applications.


\section{Use Cases of BD-RIS in 6G IoT Networks}
The integration of BD-RIS into 6G wireless networks promises to revolutionize various communication paradigms, enabling efficient, high-performance, and adaptable networks. In this section, we explore how BD-RIS can be leveraged in diverse 6G IoT applications, including smart agriculture, Industry 5.0, smart cities, and NTNs-based IoT systems, as illustrated in Fig. \ref{UseCases}.

\subsection{BD-RIS in Smart Agriculture}
Smart agriculture integrates advanced technologies such as IoT, artificial intelligence, and automation to enhance farming efficiency, productivity, and sustainability. By leveraging real-time data, farmers can optimize resource allocation, improve crop yields, and reduce environmental impact. A key enabler of smart agriculture is the deployment of dense IoT sensor networks to monitor environmental conditions, automate irrigation, and optimize resource utilization. However, rural and remote agricultural areas often suffer from poor wireless coverage, high path loss, and signal degradation caused by vegetation, soil moisture variations, and dynamic environmental conditions. Additionally, energy efficiency remains a major concern, as most IoT devices operate on battery power or energy harvesting. These IoT devices collect crucial agricultural data, including soil moisture, temperature, humidity, and crop health, enabling farmers to make informed decisions in real time. However, due to their limited energy capacity, IoT devices struggle to process and transmit this data to a central access point (AP). The communication between IoT devices and the AP is further hindered by deep fading and signal blockages caused by tree canopies, uneven terrain, and surrounding infrastructure. 

BD-RIS can significantly enhance wireless connectivity in smart agricultural environments by dynamically reconfiguring the wireless channel, thereby improving the signal strength of IoT devices at the AP. By leveraging a full scattering matrix, BD-RIS can perform advanced beamforming, focusing signals toward the AP and mitigating propagation losses caused by vegetation and terrain obstructions. This ensures reliable and seamless data transmission between IoT devices, drones, and the AP. Additionally, BD-RIS can optimize spectral efficiency and reduce interference among coexisting IoT devices, enabling scalable and energy-efficient agricultural IoT deployments.
\subsection{BD-RIS in Industry 5.0 and Beyond}
Industry 5.0 envisions a new era of intelligent, human-centric, and highly interconnected industrial environments where humans, robots, and IoT devices collaborate seamlessly in real time. Unlike Industry 4.0, which primarily focuses on automation and digitalization, Industry 5.0 emphasizes human-machine synergy, sustainability, and resilience in industrial operations. This shift demands ultra-reliable low-latency communication (URLLC), massive connectivity, and real-time data exchange to support advanced applications such as collaborative robots (cobots), autonomous guided vehicles (AGVs), predictive maintenance, and remote monitoring. However, industrial facilities present significant wireless communication challenges, including severe signal attenuation due to metal surfaces, machinery-induced multipath fading, blockages from large equipment, and electromagnetic interference from industrial processes. These factors degrade signal quality, limiting network reliability and increasing latency, which can be detrimental to critical industrial applications. 

BD-RIS can enhance industrial IoT communication by smartly reconfiguring the wireless environment, steering signals around obstacles, and mitigating interference from industrial machinery. Its ability to control signal propagation significantly improves coverage, reliability, and network adaptability in harsh industrial settings. Moreover, BD-RIS can optimize communication for cobots and AGVs by ensuring robust, low-latency wireless links, which are crucial for real-time automation, industrial safety, and predictive maintenance. BD-RIS can enhance spectral efficiency in highly congested environments, facilitating seamless interaction between machines and human operators.
Furthermore, BD-RIS can play a vital role in energy-efficient industrial networks by reducing transmission power requirements while maintaining strong connectivity. This is particularly beneficial for battery-powered industrial IoT devices and wireless sensors deployed in smart factories. As industries move toward AI-driven automation, BD-RIS will serve as a key enabler of resilient and adaptive wireless communication, ensuring seamless integration of next-generation industrial technologies.
\subsection{BD-RIS in Smart City}
Smart cities use advanced digital infrastructure and massive IoT deployments to enhance urban living, improve sustainability, and optimize public services. By integrating intelligent transportation systems, environmental monitoring, and public safety solutions, smart cities enable real time decision-making and automation to address urban challenges. The foundation of a smart city relies on ubiquitous connectivity to support applications such as autonomous vehicles, traffic management, air quality monitoring, smart surveillance, and energy-efficient infrastructure. However, ensuring seamless wireless communication in such densely populated and dynamic environments presents significant challenges. Dense urban landscapes suffer from severe signal blockages due to high-rise buildings, excessive multipath interference, and increased spectrum congestion from a high density of wireless devices. These factors can degrade network performance, leading to unreliable connectivity, high latency, and inefficient spectrum utilization. Moreover, as smart city applications demand URLLC, conventional network solutions struggle to meet these stringent requirements. 

BD-RIS is a transformative technology that actively shapes the wireless environment to increase connection and address urban challenges with communication. When strategically implemented on urban infrastructure like buildings, lampposts, and traffic lights, BD-RIS can improve coverage, eliminate signal jams, and increase spectrum efficiency. BD-RIS enables smooth V2X connection, providing dependable data sharing between cars, pedestrians, and roadside units (RSUs). Furthermore, BD-RIS improves the performance of intelligent transportation systems, smart surveillance, and public safety networks by reducing interference and allowing for effective spectrum reuse. Furthermore, BD-RIS promotes energy-efficient urban communication by optimizing power distribution and reducing unwanted signal reflections. This capacity is critical for sustainable smart city operations, as it reduces energy usage while maintaining strong network performance. 

\subsection{BD-RIS in NTNs-based IoT}
NTNs will play a pivotal role in extending IoT connectivity to remote, disaster-prone, and underserved areas where traditional terrestrial networks are impractical or economically unfeasible. NTNs comprises diverse platforms, including low Earth orbit (LEO) satellites, high-altitude platform stations (HAPS), and unmanned aerial vehicles (UAVs), which collectively provide global coverage and resilient communication infrastructures. These platforms support critical applications such as environmental monitoring, disaster response, maritime surveillance, and remote sensing, ensuring uninterrupted IoT connectivity even in the most challenging environments. Despite their advantages, NTNs-IoT networks face several fundamental challenges that limit their performance. High free-space path loss significantly weakens received signals, particularly over long distances. The mobility of LEO satellites and UAVs introduces Doppler shifts, leading to signal misalignment and synchronization issues. Additionally, constrained spectrum resources exacerbate co-channel interference, making it difficult to maintain reliable, low-latency communication between terrestrial and non-terrestrial nodes. Moreover, IoT devices with tiny aperture antennas struggle to receive satellite signals, resulting in a severely degraded link budget. 

BD-RIS emerged as an innovative solution to enhance NTNs-IoT communication by compensating for propagation losses and optimizing signal redirection, thereby improving the link budget of IoT devices. BD-RIS enables advanced beamforming techniques to dynamically adjust signal pathways, ensuring seamless handovers between LEO satellites, UAVs, and terrestrial IoT devices. This adaptability is crucial for mitigating Doppler effects, maintaining link stability, and reducing communication outages in mobile NTNs environments. In disaster response scenarios, BD-RIS can establish resilient communication links between rescue teams, autonomous drones, and satellites, ensuring real-time data transmission in emergency situations. Similarly, for maritime surveillance and environmental monitoring, BD-RIS enhances signal reception over vast oceanic and remote terrains, improving data acquisition accuracy.

\section{Case Study: BD-RIS Assisted V2V Communication}
V2V communication plays a crucial role in enabling low-latency, high-reliability data exchange between vehicles, supporting applications such as autonomous driving, traffic management, and road safety. It can be deployed as an underlay network coexisting with cellular networks, utilizing the same spectrum resources to enhance spectral efficiency. However, V2V communication faces challenges such as severe interference from neighboring vehicles, dynamic channel conditions due to high mobility, and blockages caused by urban infrastructure.

BD-RIS can effectively address these challenges by intelligently reconfiguring the wireless environment, optimizing signal propagation, and mitigating interference. By dynamically adjusting its phase shifts and amplitudes, BD-RIS can establish robust V2V links, improving signal strength and suppressing inter-vehicle interference. This is particularly beneficial in urban scenarios with dense traffic, where vehicles frequently experience signal degradation due to obstacles such as buildings and other vehicles. Additionally, in highway scenarios, BD-RIS can extend coverage and enhance communication reliability for fast-moving vehicles, ensuring seamless connectivity.

To analyze the impact of BD-RIS on V2V communication, we consider a scenario where V2V links operate as an underlay network within a cellular system, sharing the same spectrum resources. The communication between the transmitting vehicle (V2V Tx) and the receiving vehicle (V2V Rx) occurs through both direct and BD-RIS assisted links. A BD-RIS is strategically deployed between V2V Tx and V2V Rx to enhance the communication performance. The objective is to maximize the spectral efficiency of V2V communication while ensuring the quality of service (QoS) of the cellular network. This can be achieved by jointly optimizing the transmit power of V2V Tx and the phase shift design of BD-RS. The spectral efficiency maximization problem is formulated as non-convex due to interference term in rate expression, coupled variables, and BD-RIS constraint. To make the problem tractable and reduce the complexity, we transform the original problem and adopt alternating optimization technique to obtain an efficient solution.

\begin{figure}[!t]
\centering
\includegraphics[width=0.5\textwidth]{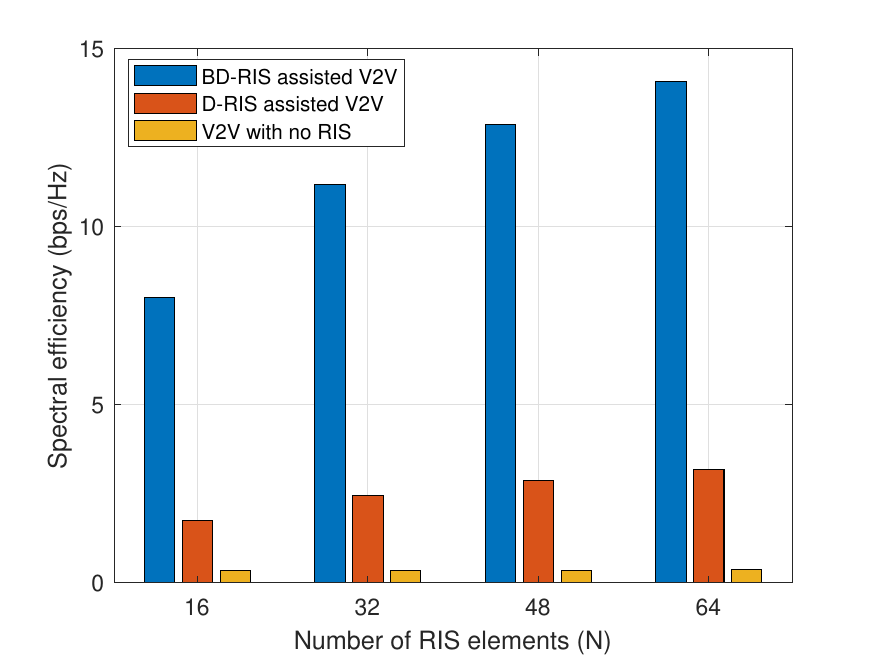}
\caption{Achievable spectral efficiency of V2V communication versus varying number of reconfigurable elements, considering V2V transmit power is 1W and cellular RSU power is 10W.}
\label{V2V}
\end{figure}

For the simulations, we assume a carrier frequency of 3.5 GHz, a cellular RSU transmit power of 10 W, a V2V transmit power of 1 W, a noise variance of 0.0001, and a varying number of BD-RIS elements ranging from 16 to 64. Fig.~\ref{V2V} illustrates the achievable spectral efficiency of V2V communication versus the number of reconfigurable elements for three scenarios: BD-RIS assisted V2V, D-RIS assisted V2V, and conventional V2V communication with no RIS. The results demonstrate that BD-RIS assisted V2V communication significantly outperforms both D-RIS assisted V2V and conventional V2V systems across all element configurations. As the number of elements increases, the spectral efficiency of V2V improves considerably due to the enhanced beamforming capability of BD-RIS. In contrast, D-RIS assisted V2V communication achieves only moderate improvements, while conventional V2V communication remains unaffected due to the absence of RIS. The superior performance of BD-RIS is attributed to its ability to efficiently control both diagonal and non-diagonal phase shifts, optimizing the wireless channel and enhancing received signal strength. These results validate the effectiveness of BD-RIS in enhancing spectral efficiency and reliability in V2V communication underlaying cellular networks.

\begin{figure}[!t]
\centering
\includegraphics [width=0.48\textwidth]{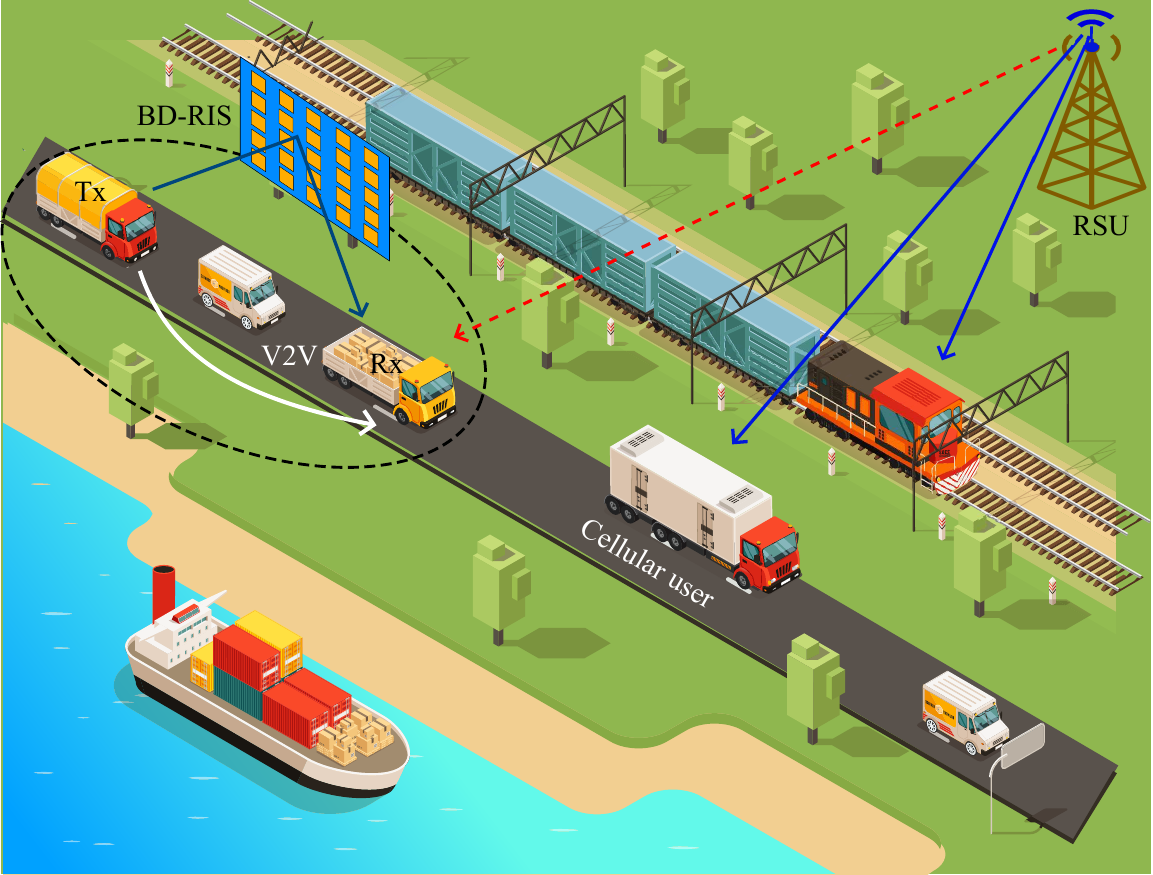}
\caption{BD-RIS assisted V2V communication underlay cellular network.}
\label{D2Dsm}
\end{figure}

\section{Research Challenges and Future Directions}

Despite the significant benefits BD-RIS offers for future 6G wireless and IoT networks, several challenges must be addressed for its full potential to be realized. Current research has only begun to explore these hurdles.

\subsection{Current Challenges}

\subsubsection{BD-RIS Hardware Design}
BD-RIS hardware development encounters significant obstacles, primarily due to the intricate design of reconfigurable impedance networks and the need for large-scale integration of controllable elements. As the array size grows, so does the complexity and cost, particularly when design considerations include specialized elements (e.g., back-to-back patch structures or narrow-beam configurations). Additionally, the coordination required to manage a large number of tunable elements further amplifies engineering challenges. Effective solutions to these challenges are still under exploration and development.

\subsubsection{Adaptive Channel Estimation}
Accurate channel estimation poses a significant challenge for BD-RIS-enabled systems, given the inherent complexity of controlling both diagonal and non-diagonal elements. Although semi-passive or on-off pilots can be sufficient for diagonal RIS configurations, these approaches typically fail to capture the full wavefront manipulation introduced by BD-RIS. In terrestrial deployments, multi-path fading and dynamic interference patterns make it difficult to disentangle individual channel components, necessitating more sophisticated parameter extraction and estimation techniques. In non-terrestrial networks (NTNs), the problem is further compounded by satellite mobility, Doppler shifts, and varying atmospheric conditions. Moreover, IoT scenarios often require energy-efficient estimation in dense or large-scale environments, where battery-powered devices operate over extended lifespans. Smart agriculture, for instance, spans vast areas with potentially weak signal returns, making it imperative to adopt estimation schemes that minimize overhead while ensuring sufficient accuracy. 
 
\subsubsection{Non-Ideal Hardware Effects}
Real-world implementations of BD-RIS inevitably suffer from hardware imperfections, such as amplifier nonlinearities, phase noise, and mutual coupling among neighboring elements, all of which can degrade beamforming accuracy. In terrestrial systems, multi-path propagation and shadowing further magnify these impairments, whereas non-terrestrial networks (NTNs) face additional challenges from satellite mobility and atmospheric variations. Effective mitigation in both environments demands adaptive calibration routines and compensation algorithms, often requiring advanced signal processing or machine learning techniques. Moreover, IoT networks commonly adopt discrete-value BD-RIS elements—particularly in hybrid modes—to simplify hardware and reduce power consumption. However, the quantization of amplitude and phase in these discrete implementations creates inter-element dependencies that complicate wavefront shaping and constraint satisfaction, underscoring the critical need for robust hardware-aware designs.

\subsubsection{Signal Detection Sensitivity}
Accurate signal detection in IoT networks becomes especially challenging when transmission distances are long, interference is prevalent, and nodes operate under strict power constraints. These factors demand that BD-RIS maintain high sensitivity despite potential noise, fading, and complex interference environments. Large-scale, dense deployments—common in IoT applications such as smart cities or agricultural monitoring—further compound the difficulty of distinguishing weak IoT signals from background interference. Advanced signal processing techniques, including ML-based noise differentiation and adaptive thresholding, can help mitigate these effects by refining detection algorithms in real-time. By intelligently shaping the wavefront and compensating for dynamic channel conditions, BD-RIS can bolster detection performance, ensuring robust connectivity even in harsh or remote IoT scenarios.

\subsection{Future Research Directions}

\subsubsection{The Convergence of AI/ML}
 AI/ML with BD-RIS facilitates autonomous, real-time tuning of reflection coefficients to meet dynamic channel and user requirements. By employing data-driven methods such as reinforcement learning and deep neural networks, BD-RIS can effectively manage complex tasks like interference cancellation, beamforming, and power allocation under heterogeneous IoT conditions. In non-terrestrial networks, these techniques can predict mobility and Doppler shifts for satellites or high-altitude platforms, ensuring robust links and effective spectrum management. Collectively, this AI-driven approach enhances network reliability, reduces latency, and enables scalable deployments across various 6G IoT use cases.

\subsubsection{Joint Communications and Sensing}
 Joint Communication and Sensing (JCAS) leverages a shared radio resource to support both data transmission and environmental monitoring, and BD-RIS can significantly enhance these capabilities by precisely shaping and steering the signal wavefront. By adjusting amplitude and phase across its reconfigurable elements, BD-RIS reduces mutual interference between communication and sensing functions, improving target detection and localization accuracy. For IoT networks, particularly in large-scale or hard-to-reach environments like smart agriculture, BD-RIS-enabled JCAS can offer seamless coverage for autonomous monitoring (e.g., soil moisture sensing) while simultaneously transmitting crucial data in real-time. Moreover, in NTNs—where environmental conditions change rapidly—BD-RIS can dynamically reconfigure beam patterns to maintain optimal sensing performance, compensating for factors such as high mobility and challenging propagation paths. This adaptability ultimately maximizes spectrum efficiency and system reliability, enabling future 6G IoT networks to integrate communication and sensing services without compromising on performance or energy constraints.

\subsubsection{Physical Layer Security}
Physical Layer Security is critical in future wireless networks where massive IoT deployments and non-terrestrial infrastructures introduce new vulnerabilities. BD-RIS can dynamically tailor the propagation environment, directing signals to legitimate receivers while minimizing leakage to unintended parties. By controlling both the amplitude and phase of scattered waves, BD-RIS can create steep nulls or destructive interference in the eavesdropper’s direction, significantly reducing the probability of interception. This capability becomes even more valuable in IoT networks that transmit sensitive data—such as financial transactions or healthcare information—and in NTNs where high mobility and variable link conditions can be exploited by adversaries. Additionally, BD-RIS-assisted systems can enhance the structure of the artificial noise for steering patterns to confuse potential eavesdroppers or jammers, effectively boosting secrecy capacity and robustness against active attacks. These adaptive security measures complement traditional cryptographic techniques by adding an extra layer of physical layer protection, making 6G IoT and NTNs networks both more resilient and more secure.

\section{Conclusion}
In this paper, we presented BD-RIS as a powerful extension of conventional diagonal RIS, highlighting its ability to simultaneously manage phase and amplitude for enhanced wavefront control. We explored various 6G IoT scenarios—ranging from smart agriculture and Industry 5.0 to NTNs—where BD-RIS can offer significant benefits in terms of coverage, interference mitigation, and energy efficiency. Our V2V use case further demonstrated its potential for boosting spectral efficiency under realistic network conditions. Despite these advantages, challenges remain in hardware design, channel estimation, and handling non-ideal effects, emphasizing the need for continued research. Looking ahead, integrating AI/ML-based optimization, JCAS functionalities, and robust physical layer security stands out as a promising direction to fully realize BD-RIS’s potential for next-generation 6G IoT systems.

\ifCLASSOPTIONcaptionsoff
  \newpage
\fi

\bibliographystyle{IEEEtran}
\bibliography{Wali_EE}

\end{document}